\documentclass[aps,preprint,nofootinbib,amsmath]{revtex4}
\usepackage{amsmath}
\usepackage{graphicx}
\usepackage{amssymb}

\makeatletter

\begin{document}
\newcommand{\met}{\not\!\! E_{T}}

\begin{flushright}UCRHEP-T432\\
MSUHEP-070702\par\end{flushright}

\title{Signatures of Extra Gauge Bosons in the Littlest Higgs Model with
T-parity at Future Colliders}

\author{Qing-Hong Cao}

\email{qcao@ucr.edu}

\affiliation{Department of Physics and Astronomy, University of California at
Riverside, Riverside, CA 92321}

\author{Chuan-Ren Chen}

\email{crchen@pa.msu.edu}

\affiliation{Department of Physics and Astronomy, Michigan State University, East
Lansing, MI 48824}

\begin{abstract}
We study the collider signatures of a T-odd gauge boson $W_{H}$ pair
production in the Littlest Higgs Model with T-parity (LHT) at Large
Hadron Collider (LHC) and Linear Collider (LC). At the LHC, we search
for the $W_{H}$ boson using its leptonic decay, i.e. $pp\to W_{H}^{+}W_{H}^{-}\to A_{H}A_{H}\ell^{+}\nu_{\ell}\ell^{\prime-}\bar{\nu}_{\ell^{\prime}}$,
which gives rise to a collider signature of $\ell^{+}\ell^{\prime-}+\met$.
We demonstrate that the LHC not only has a great potential of discovering
the $W_{H}$ boson in this channel, but also can probe enormous parameter
space of the LHT. Due to four missing particles in the final state,
one cannot reconstruct the mass of $W_{H}$ at the LHC. But such a
mass measurement can be easily achieved at the LC in the process of
$e^{+}e^{-}\to W_{H}^{+}W_{H}^{-}\to A_{H}A_{H}W^{+}W^{-}\to A_{H}A_{H}jjjj$.
We present an algorithm of measuring the mass and spin of the $W_{H}$
boson at the LC. Furthermore, we illustrate that the spin correlation
between the $W$ boson and its mother particle ($W_{H}$) can be used
to distinguish the LHT from other new physics models.
\end{abstract}
\maketitle

\section{introduction}

It has been shown that the collective symmetry breaking mechanism
implemented in Little Higgs models~\cite{Arkani-Hamed:2001nc} provides
an interesting solution to the {}``little hierarchy problem'' (also
see~\cite{Schmaltz:2005ky,Perelstein:2005ka} for recent review).
The Littlest Higgs model, a $SU(5)/SO(5)$ nonlinear sigma model proposed
in Ref.\ \cite{Arkani-Hamed:2002qy}, is one of the most economical
and interesting models discussed in the literature. In the Littlest
Higgs Model, the global symmetry $SU(5)$ is broken down to $SO(5)$
by a $5\times5$ symmetric tensor at the scale $f$. Simultaneously,
the gauged $[SU(2)\times U(1)]_{1}\times[SU(2)\times U(1)]_{2}$,
a subgroup of $SU(5)$, is broken to the diagonal $SU(2)_{W}\times U(1)_{Y}$,
a subgroup of $SO(5)$. A vector-like quark, $T_{+}$, is introduced
in the top sector to cancel the quadratic divergence contribution
to Higgs boson mass from the Standard Model (SM) top quark loop. The
low energy electroweak precision tests (EWPT), however, enforce the
symmetry breaking scale $f$ to be larger than about $4\,{\rm TeV}$.
As a result, the cut-off scale $\Lambda\sim4\pi f$ becomes so large
that the fine tuning between the cut-off scale and the electroweak
scale is needed again~\cite{Csaki:2002qg,Csaki:2003si,Hewett:2002px,Chen:2003fm,Kilian:2003xt,Han:2004az}.
The Littlest Higgs model with T-parity (LHT)~\cite{Cheng:2003ju,Cheng:2004yc,Low:2004xc}
was proposed by imposing a discrete $Z_{2}$ symmetry, called T-parity,
into the Littlest Higgs model. T-parity~\cite{Cheng:2003ju,Cheng:2004yc,Low:2004xc}
is a symmetry which exchanges the gauge boson fields of the two gauged
$SU(2)\times U(1)$ groups, i.e. $[SU(2)\times U(1)]_{1}\leftrightarrow[SU(2)\times U(1)]_{2}$.
One direct consequence of the T-parity is the absence of the mixing
between the extra heavy gauge bosons and the SM gauge bosons, because
they have different T-parity quantum numbers. The constraints from
EWPT are alleviated so that the scale $f$ could be as low as $500\,{\rm GeV}$~\cite{Hubisz:2005tx}.

In order to incorporate the T-parity systematically, extra fermion
fields have to be introduced. One needs two sets of gauge boson fields
and fermion fields transforming independently under $[SU(2)\times U(1)]_{1,\,2}$.
One of the two possible linear combinations of the fields from two
different sets is assigned to be the SM field and another combination
is the extra heavy field. The heavy particles (except the vector-like
$T_{+}$) are odd under the T-parity while the SM particles are even.
With the exact T-parity embedded, the effective operators which mix
T-odd and T-even fields are absent. Details of the LHT considered
in this paper have been shown in Refs.~\cite{Hubisz:2004ft,Belyaev:2006jh}.
Here, we only layout the mass spectrum of the particles relevant to
our study, which are $A_{H}$ (T-parity partner of photon), $W_{H}$
(T-parity partner of $W$ boson), $\ell_{-}$ (T-odd lepton) and $q_{-}$
(T-odd quark) ,\begin{eqnarray*}
m_{A_{H}} & \simeq & \frac{g^{\prime}f}{\sqrt{5}},\,\, m_{W_{H}}\simeq gf,\,\, m_{\ell_{-}}\simeq\sqrt{2}\kappa_{\ell}f\,,\,\, m_{q_{-}}\simeq\sqrt{2}\kappa_{q}f,\end{eqnarray*}
 where $g^{\prime}$ and $g$ are the hypercharge and weak gauge coupling,
respectively, and $\kappa_{\ell}$ ($\kappa_{q}$) is the Yukawa type
coupling introduced in the interaction which generates the T-odd lepton
(quark) mass. $A_{H}$ is usually the lightest T-odd particle (LTP)
which cannot further decay into the SM particles and thus plays as
the dark matter candidate. With the allowed low mass scale, these
extra T-odd particles have significant impacts on the phenomenology~\cite{Chen:2006cs,Choudhury:2006xa,Blanke:2006eb,Hundi:2006rh,Cao:2006wk,Blanke:2006sb,Blanke:2006xr,Blanke:2007db,Blanke:2007ee,Blanke:2007wr,Choudhury:2006sq,Yue:2006hn,Chen:2006ie,Hong-Sheng:2007ve,Kai:2007ji,Wang:2007zx,Yue:2007ww}.
Large Hadron Collider (LHC) at CERN has a great potential to copiously
produce these new particles. Some studies about collider phenomenology
of the LHT have been presented recently~\cite{Hubisz:2004ft,Freitas:2006vy,Belyaev:2006jh,Matsumoto:2006ws,Choudhury:2006mp,Carena:2006jx,Cao:2006wk}. 

Current EWPT only impose constraints on the parameter space of the
LHT. Due to the T-parity, the new T-odd particles have to be produced
in pairs at the colliders. The fact that at least two missing particles
remain in the final state makes it difficult to measure the model
parameters of the LHT, see details in the discussions of the LHC phenomenology.
In order to test the LHT at the LHC, one has to observe the new physics
signatures in various independent channels. By comparing the model
parameters extracted out from those channels one might be able to
check the consistency of the LHT. For that, the $W_{H}^{+}W_{H}^{-}$
production is of importance because the mass of the heavy gauge boson
$W_{H}$ ($m_{W_{H}}$) depends on $f$ only. One thus can directly
determine the symmetry breaking scale $f$ from the $W_{H}$ mass
measurement\ %
\footnote{Recently, Ref.~\cite{Cao:2006wk} proposed that one can measure $f$
using the spin correlation between the top quark pair in the process
of $pp\to T_{-}T_{-}\to tA_{H}\bar{t}A_{H}$, where $T_{-}$ is the
T-parity partner of the vector-like $T_{+}$.%
}. In this paper, we examine the discovery potential of the $W_{H}^{+}W_{H}^{-}$
pair production at the LHC and present a strategy of measuring the
mass and spin of $W_{H}$ at the LC. The matrix elements of both signal
and background processes are calculated using MadGraph\ \cite{Stelzer:1994ta,Maltoni:2002qb}
while the widths of the new T-odd particles are calculated in CalcHEP\ \cite{Pukhov:2004ca}
with the model file given by Ref.\ \cite{Belyaev:2006jh}. Agreement
of both programs at the level of new gauge boson production has been
checked. The rest of this paper is organized as follows. In Sec.~\ref{sec:rate-ps},
we present the cross sections of the $W_{H}^{+}W_{H}^{-}$ pair production
at the LHC and at the LC. We also discuss the decay pattern of $W_{H}$
and present the unitarity constraints on the parameter space of the
LHT from effective four fermion interaction operators. The collider
phenomenology of the LHC and the LC is shown in Sec.~\ref{sec:Collider-lhc}
and Sec.\ \ref{sec:Collider-ilc}, respectively. Finally, we conclude
in Sec.~\ref{sec:Conclusion}.

\section{production and decay of $W_{H}$ boson~\label{sec:rate-ps}}

\begin{figure}
\includegraphics[scale=0.5]{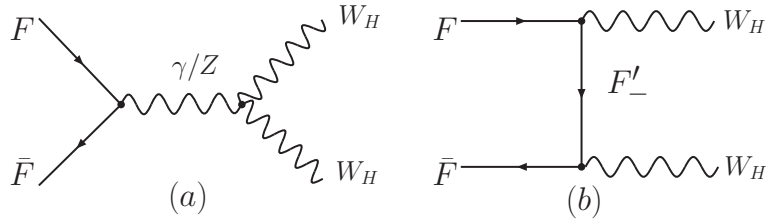}

\caption{Feynman diagrams for a $W_{H}$ pair production.\label{fig:whprodlhc}}
\end{figure}

The tree-level diagrams for a $W_{H}$ pair production are shown in
Fig.\ \ref{fig:whprodlhc}, where $F$ and $F^{\prime}$ denote the
quarks at the LHC while the electron and electron-neutrino at the
LC. The $W_{H}$ boson pair can be produced either via the $s$-channel
process with the photon and $Z$ boson exchanged or via the $t$-channel
process with a T-odd fermion exchanged. Since the $t$-channel diagram
involves the heavy T-odd fermion, its contribution depends on both
$m_{W_{H}}$ and $m_{F_{-}}$. In this work we choose the model parameters
($f$,\ $\kappa_{q/\ell}$) instead of the physical masses of the
new particles as the theoretical inputs.

\subsection{$W_{H}$ production at the LHC}

\begin{figure}
\includegraphics[clip,scale=0.6]{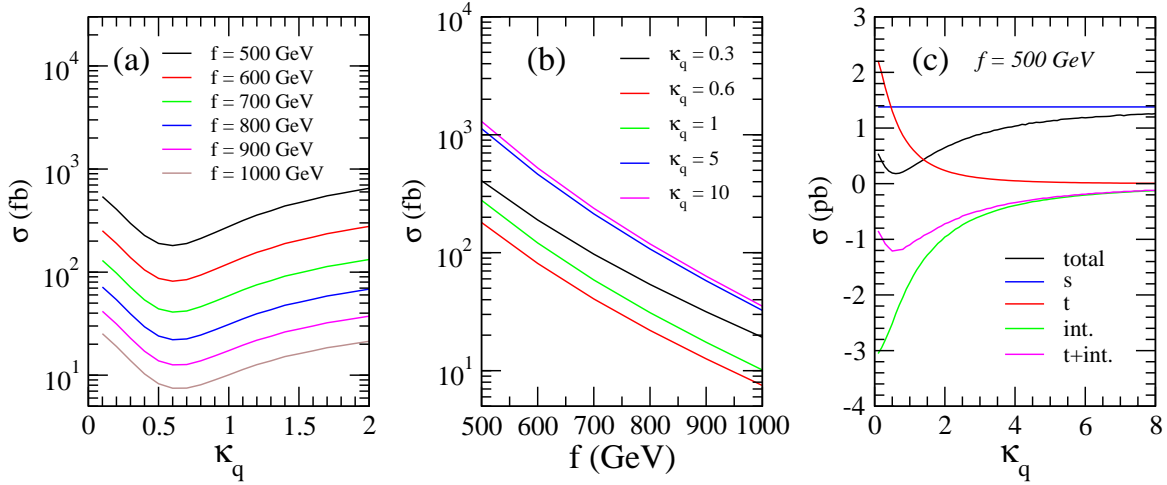}

\caption{The total cross section of a $W_{H}^{+}W_{H}^{-}$ pair production
at the LHC for various parameters $f$ and $\kappa_{q}$.~\label{fig:xsec-k} }
\end{figure}

In Fig.\ \ref{fig:xsec-k}(a)\ and\ \ref{fig:xsec-k}(b) we show
the total cross section of the $W_{H}$ pair production as a function
of $\kappa_{q}$ and $f$, respectively. The T-odd quark in the $t$-channel
diagram affects the total cross section significantly: (i) for $500\,{\rm GeV}<f<1000\,{\rm GeV}$,
there exists a $\kappa_{q}^{min}$($\sim0.6$) which minimizes the
total cross section; (ii) for a fixed $\kappa_{q}$, the cross section
decreases rapidly with increasing $f$. In order to understand why
the minimum of the total cross section occurs, we separate the total
cross section into three pieces, \begin{equation}
\sigma_{tot}=\sigma_{s}+\sigma_{t}+\sigma_{{\rm int}},\label{eq:tot-xsec}\end{equation}
where $\sigma_{s}$, $\sigma_{t}$ and $\sigma_{{\rm int}}$ denote
the contributions of the $s$-channel diagram, $t$-channel diagram
and the interference between the $s$- and $t$-channel diagrams,
respectively. For illustration, we choose $f=500\,{\rm GeV}$ and
plot each individual contribution in Fig.\ \ref{fig:xsec-k}(c).
The $s$-channel diagram involves the gauge bosons only, therefore,
its contribution depends on $f$ but not on $\kappa_{q}$, cf. the
flat blue curve. On the contrary, the $t$-channel contribution decreases
with increasing $\kappa_{q}$, because the mass of the T-odd quark
in the $t$-channel propagator grows with increasing $\kappa_{q}$,
cf. the red curve. Although the $s$-channel and $t$-channel contributions
are both constructive, their interference is destructive. The total
cross section reaches the minimum when $\kappa_{q}\sim\kappa_{q}^{min}$,
where the $s$- and $t$-channel contributions are comparable. When
$\kappa_{q}>\kappa_{q}^{min}$, the total cross section is dominated
by the $s$-channel contribution, therefore it drops rapidly with
increasing $f$ since the $s$-channel contribution suffers from the
$1/\hat{s}$ suppression ( $\hat{s}$ is the invariant mass of the
$W_{H}$ boson pair). When $\kappa_{q}\gg\kappa_{q}^{min}$, the total
cross section approaches to the $s$-channel contribution and both
the $t$-channel contribution and the interference effect are negligible. 

\subsection{$W_{H}$ production at the LC}

\begin{figure}
\includegraphics[clip,scale=0.6]{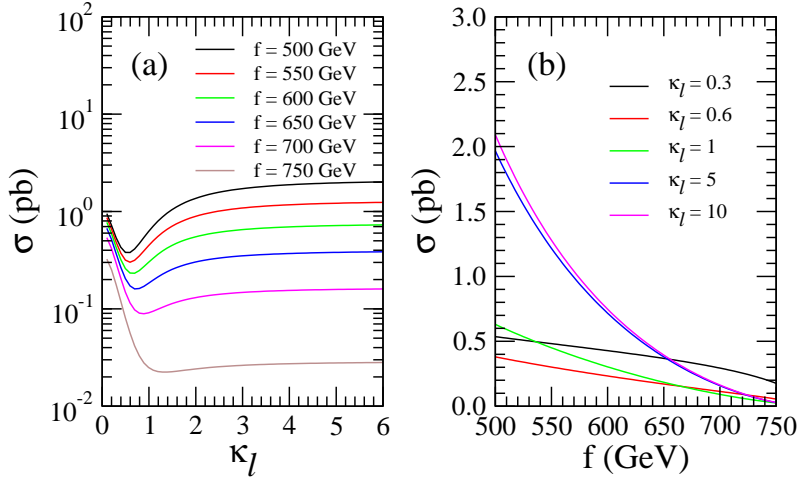}

\caption{Total cross section of a $W_{H}^{+}W_{H}^{-}$ pair production at
the LC for various $f$ and $\kappa_{\ell}$.~\label{fig:xsec-ILC}}
\end{figure}

We present the total cross section of the $W_{H}$ pair production
at the LC as a function of $\kappa_{\ell}$ and $f$ in Fig.\ \ref{fig:xsec-ILC}(a)
and (b), respectively. In analogue to the $W_{H}$ pair production
at the LHC, there also exists a $\kappa_{\ell}^{min}$ due to the
destructive interference effect, but $\kappa_{\ell}^{min}$ is very
sensitive to $f$ at the LC. As shown in Fig.\ \ref{fig:xsec-ILC}(a),
$\kappa_{\ell}^{min}$ shifts from about 0.5 to 1.0 when $f$ increases
from 500\ GeV to 750\ GeV. We also note that the total cross section
of a small $\kappa_{\ell}$, e.g. $\kappa_{\ell}=0.3$, drops much
slower than the total cross section of a large $\kappa_{\ell}$, see
Fig.~\ref{fig:xsec-ILC}(b). 

\begin{figure}
\includegraphics[clip,scale=0.6]{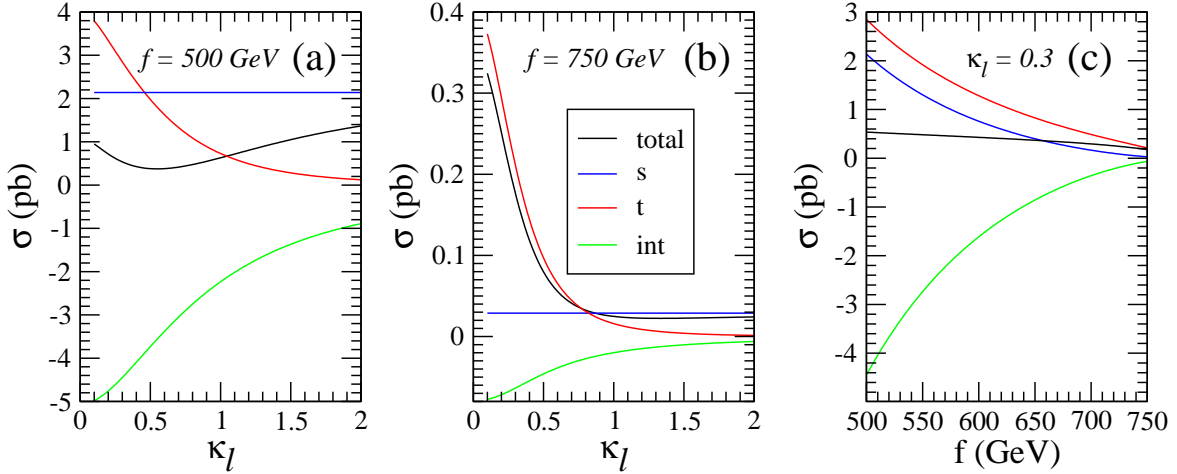}

\caption{The distributions of $s-$, $t-$channel diagrams and interference
term in the $W_{H}^{+}W_{H}^{-}$ production at the LC.~\label{fig:xsec-ILC-s-t-int}}
\end{figure}

Following the LHC study, we split the total cross section into the
$s$-channel, $t$-channel and the interference contributions. In
Fig.\ \ref{fig:xsec-ILC-s-t-int} we explicitly plot the total cross
section (black curve), the $s$-channel contribution (blue curve),
the $t$-channel contribution (red curve) and the interference contribution
(INT) (green curve). Fig.\ \ref{fig:xsec-ILC-s-t-int}(a) and (b)
show the total cross section as a function of $\kappa_{\ell}$ for
$f=500\,{\rm GeV}$ and $750\,{\rm GeV}$, respectively. We have learned
from the LHC study that the minimal cross section for a fixed $f$
occurs when $\sigma(s)\simeq\sigma(t)$. When $f$ increases from
500 GeV to 750 GeV, the $s$-channel contribution drops rapidly since
it suffers from the $1/\hat{s}$ suppression, but on the other hand,
the $t$-channel contribution does not. Of course, increasing $f$
value will increase the mass of $W_{H}$ boson and reduce the $t$-channel
contribution, but the suppression in the $t$-channel contribution
is much less than that in the $s$-channel contribution. Therefore,
the position for $\sigma(s)\simeq\sigma(t)$ is shifted to larger
$\kappa_{\ell}$ region. The reason why the cross section of $\kappa_{\ell}=0.3$
drops slowly in the large $f$ region can also be understood from
the competition between the $s$- and $t$-channel contributions.
In Fig.\ \ref{fig:xsec-ILC-s-t-int}(c) we show the total cross section
as a function of $f$ for $\kappa_{\ell}=0.3.$ For such a small $\kappa_{\ell}$,
the T-odd neutrino's mass is small ($m_{\nu-}\simeq0.42f$). Then
the $t$-channel contribution dominates over the $s$-channel contribution.
In the large $f$ region, i.e. $600\,{\rm GeV}<f<750\,{\rm GeV}$,
the $s$-channel contribution as well as the interference effect both
decrease to zero, and the total cross section approaches to the $t$-channel
contribution which does not drop rapidly with increasing $f$.

\subsection{Decay of the $W_{H}$ boson}

The $W_{H}$ boson will decay into a T-odd particle and a T-even SM
particle. Its decay pattern is mainly determined by the masses of
new T-odd particles. In the LHT,\begin{eqnarray}
m_{A_{H}}\simeq\frac{g^{\prime}f}{\sqrt{5}}\simeq0.156f, &  & m_{W_{H}}\simeq gf\simeq0.653f,\nonumber \\
m_{\ell-}\simeq\sqrt{2}\kappa_{\ell}f\simeq1.414\kappa_{\ell}f, &  & m_{q-}\simeq\sqrt{2}\kappa_{q}f\simeq1.414\kappa_{q}\, f.\label{eq:mass}\end{eqnarray}
It is clear that the $A_{H}$ boson is always lighter than the $W_{H}$
boson. But the T-odd quark (lepton) can be heavier or lighter than
the $W_{H}$ boson, depending on the parameter $\kappa_{q}$($\kappa_{\ell}$).
Let us denote $F_{-}$ as the T-odd fermion whose mass $m_{F_{-}}$
is $\sqrt{2}\kappa f$. When $\kappa<0.11$, $m_{F_{-}}<m_{A_{H}}<m_{W_{H}}$,
therefore the T-odd lepton or T-odd quark will play the role as the
dark matter candidate. As pointed out in Ref.~\cite{Primack:1988zm},
the dark matter candidates should be charge neutral and colorless
objects. Hence, we focus our attention to the case of $\kappa_{\ell}\,(\kappa_{q})>0.11$
throughout this study, i.e. demanding $A_{H}$ to be the lightest
T-odd particle. When both $\kappa_{q}$ and $\kappa_{\ell}$ are larger
than 0.462, i.e. $m_{A_{H}}<m_{W_{H}}<m_{F_{-}}$, the $W_{H}$ boson
only decays via the $W_{H}\to W+A_{H}$ channel. When $0.11<\kappa<0.462$,
i.e. $m_{A_{H}}<m_{F_{-}}<m_{W_{H}}$, then $W_{H}$ boson can decay
into either $WA_{H}$ or $F_{-}F^{\prime}$ ($F^{\prime}$ being the
usual SM fermion). 

\begin{figure}
\includegraphics[clip,scale=0.5]{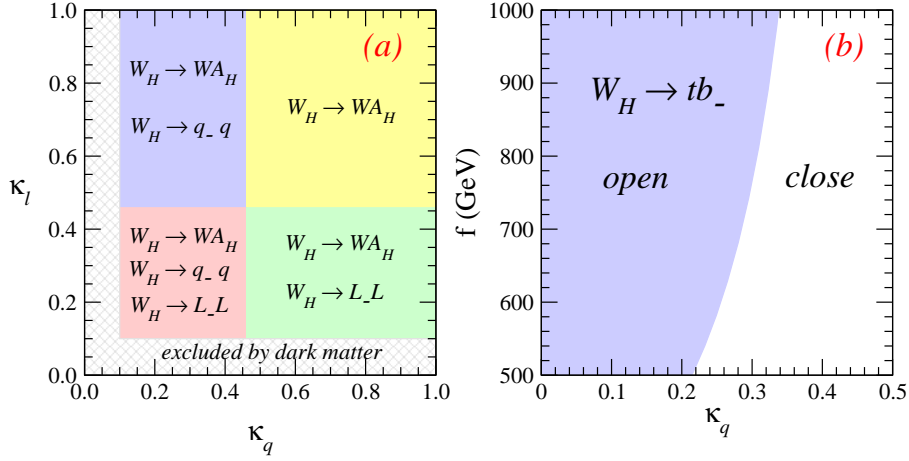}

\caption{(a) Pictorial illustration of the decay pattern of the $W_{H}$ boson
in the plane of $\kappa_{\ell}$ and $\kappa_{q}$ ; (b) allowed region
(blue) of $\kappa_{q}$ for the $W_{H}\to tb_{-}$ mode being opened.\label{fig:decay_pattern}}
\end{figure}

\begin{figure}
\includegraphics[clip,scale=0.6]{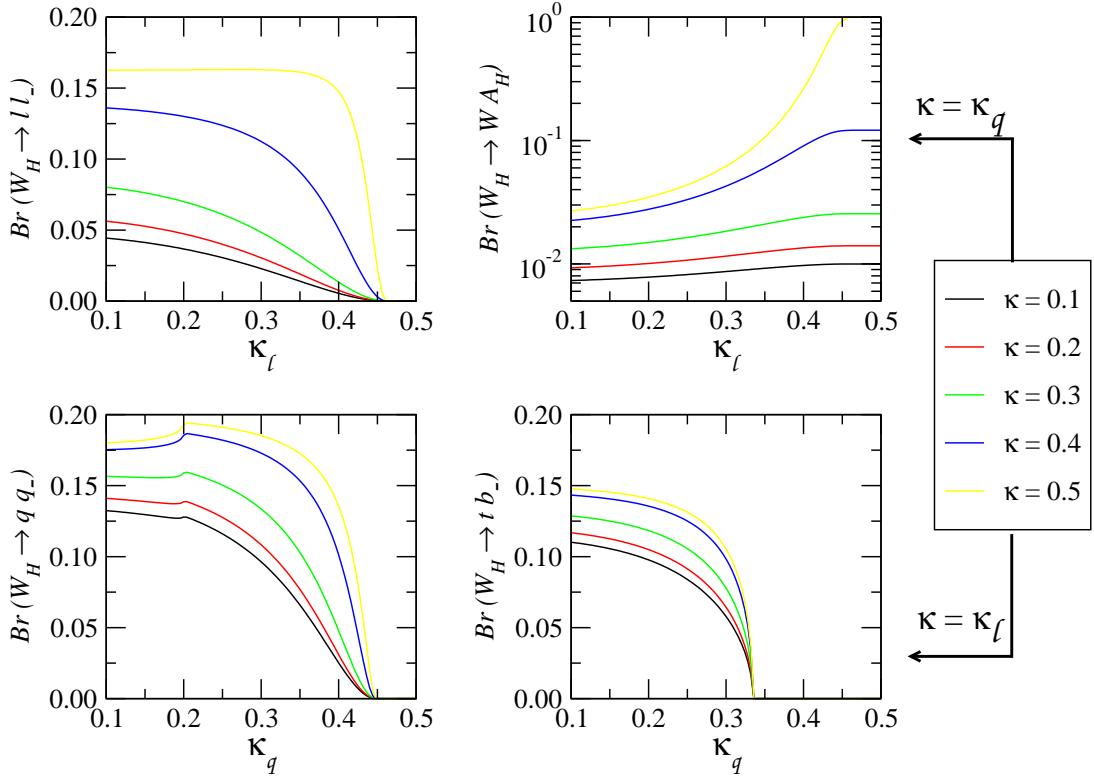}

\caption{Decay branching ratios of the $W_{H}$ boson for $f=500\,{\rm GeV}$.\label{fig:whdecay-br}}
\end{figure}

In Fig.\ \ref{fig:decay_pattern}(a) we summarize the decay pattern
of $W_{H}$ in the plane of $\kappa_{q}$ and $\kappa_{\ell}$, where
the following decay modes are considered:\begin{eqnarray}
W_{H} & \to & WA_{H}\to\ell\bar{\ell}'\left(q\bar{q}'\right)A_{H},\\
W_{H} & \to & \ell_{-}\nu_{\ell}\to\ell A_{H}\nu_{\ell},\\
W_{H} & \to & \nu_{\ell-}\ell\to\nu_{\ell}A_{H}\ell,\\
W_{H} & \to & q_{-}q'\to qA_{H}q'.\end{eqnarray}
Here, $\ell$($\nu$,$q$) denotes the charged leptons (neutrinos,
quarks). We also include the subsequent decay of the second T-odd
fermions whose decay branching ratio is 100\% for $0.11<\kappa<0.462$.
In the above decay modes, the $W_{H}\to tb_{-}\to tbA_{H}$ mode is
special because of large top quark mass ($m_{t}$). In order to open
the decay mode $W_{H}\to tb_{-}$, the mass constraint $m_{W_{H}}>m_{t}+m_{b_{-}}$
has to be satisfied and the allowed region of $\kappa_{q}$ and $f$
is shown in Fig.\ \ref{fig:decay_pattern}(b). As shown in Eq.\ (\ref{eq:mass}),
the mass relation between the $W_{H}$, $A_{H}$ and $F_{-}$ is fixed
by $\kappa$ and does not depend on $f$. Thus, the decay branching
ratios of the $W_{H}\to WA_{H}$ and $W_{H}\to F_{-}F^{\prime}$ modes
do not depend on $f$ if the $tb_{-}$ mode is not opened. Once the
$tb_{-}$ mode is opened, the decay branching ratios of other modes
will be slightly reduced. In Fig.\ \ref{fig:whdecay-br} we show
the decay branching ratios of the $W_{H}$ boson as a function of
$\kappa_{\ell}$ and $\kappa_{q}$, respectively. Explicit numbers
of the decay branching ratios for the selected benchmark points are
listed in table~\ref{tab:BR}.

\begin{table}

\caption{Decay branching ratios ($\%$) of the $W_{H}$ boson for a few benchmark
points, where $\ell=e,\mu,\tau$, $\nu=\nu_{e},\nu_{\mu},\nu_{\tau}$,
$U=u,c$ and $D=d,s$. Note that all the SM fermions (except the top
quark) are treated as massless. ~\label{tab:BR}}

\begin{tabular}{c|ccc|ccc|ccc|c}
\hline 
&
\multicolumn{3}{c|}{$\kappa_{\ell}=0.3$\ \ $\kappa_{q}=0.3$}&
\multicolumn{3}{c|}{$\kappa_{\ell}=0.5$\ \  $\kappa_{q}=0.3$}&
\multicolumn{3}{c|}{$\kappa_{\ell}=0.3$\ \ $\kappa_{q}=0.5$}&
\multicolumn{1}{c}{$\kappa_{\ell}=0.5$\ \ $\kappa_{q}=0.5$}\tabularnewline
\hline 
$f$(GeV)&
\ \ \ 500\ \ \ &
\ \ \ 700\ \ \ &
\ 1000&
\ \ \ 500\ \ \ &
\ \ \ 700\ \ \ &
\ 1000&
\ \ \ 500\ \ \ &
\ \ \ 700\ \ \ &
\ 1000&
$>500$ \tabularnewline
\hline 
$\ell_{-}\nu$&
4.45&
4.61&
4.33&
0&
0&
0&
15.0&
15.9&
16.3&
0\tabularnewline
$\nu_{-}\ell$&
4.84&
4.81&
4.41&
0&
0&
0&
16.3&
16.5&
16.6&
0\tabularnewline
\hline
$U_{-}D$&
14.5&
14.4&
13.2&
20.1&
20.1&
17.9&
0&
0&
0&
0\tabularnewline
$D_{-}U$&
13.4&
13.8&
13.0&
18.5&
19.3&
17.6&
0&
0&
0&
0\tabularnewline
$t_{-}b$&
14.5&
14.4&
13.2&
20.1&
20.1&
17.9&
0&
0&
0&
0\tabularnewline
$tb_{-}$&
0&
0&
7.79&
0&
0&
10.6&
0&
0&
0&
0\tabularnewline
\hline
$WA_{H}$&
1.84&
0.8&
0.33&
2.55&
1.12&
0.45&
6.19&
2.76&
1.25&
100\tabularnewline
\hline
\end{tabular}
\end{table}

\subsection{Unitarity constraints on $\kappa_{q}$ and $\kappa_{\ell}$}

Let us examine the low energy constraints on $\kappa_{\ell}$ and
$\kappa_{q}$ in this section before studying the phenomenology of
the $W_{H}$ boson. The mass constraints on T-odd fermion, i.e. the
lepton ($\ell_{-}$) and quark ($q_{-}$), could be derived from four-fermion
interaction operators $O(ffff)$. 

\begin{figure}
\includegraphics[clip,scale=0.4]{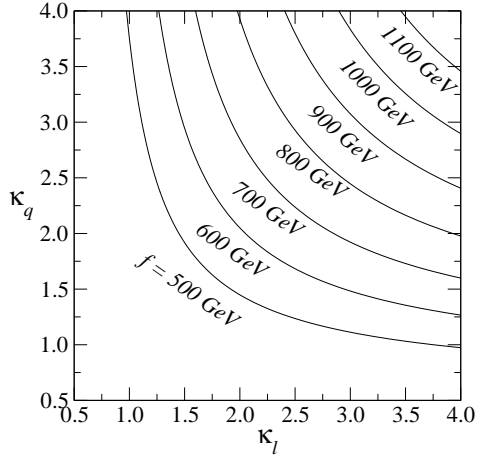}

\caption{Allowed region of $\kappa_{\ell}$ and $\kappa_{q}$ for various
values of $f$. The region below each curve is allowed. \label{fig:allowed-k}}
\end{figure}

The most general chirally invariant form of the four fermion interaction
reads\[
\frac{g^{2}}{2\Lambda^{2}}\bar{\psi}_{L}\gamma^{\mu}\psi_{L}\bar{\psi}_{L}\gamma_{\mu}\psi_{L},\]
where $\Lambda$ is the new physics scale. One then can determine
the scale $\Lambda$ unambiguously from the unitarity condition by
setting $g^{2}(\Lambda)/4\pi=1$ for the new strong interaction coupling.
For example, $\Lambda(eeee)\,>10.3\,{\rm TeV}$, $\Lambda(eedd)\,>26.4\,{\rm TeV},$
and $\Lambda(uudd)\,>2.4\,{\rm TeV}$ at $95\%$ confidence level~\cite{Yao:2006px}.
Using these limits, we can calculate the upper bound on T-odd fermion
masses. If we assume the universal mass for T-odd lepton ($\ell_{-}$)
and quark ($q_{-}$), i.e. $\kappa_{\ell}=\kappa_{q}=\kappa$, the
strongest constraint is from $O(eedd)$~\cite{Hubisz:2005tx}, which
leads to\begin{equation}
\kappa_{\ell}=\kappa_{q}\leq3.4\frac{f}{{\rm TeV}}\,.\label{eq:m_Todd_uni}\end{equation}
However, there is no physics reason to believe that the lepton and
quark sectors will share the same $\kappa$. In this work we will
treat $\kappa_{\ell}$ and $\kappa_{q}$ separately. As a result,
the masses of the T-odd leptons differ from the masses of the T-odd
quarks. In order to avoid the problem of flavor changing neutral current
(FCNC), we further assume $\kappa_{\ell}$ and $\kappa_{q}$ are universal
individually and also diagonal in the flavor space. Under this assumption,
we obtain the constraints on $\kappa_{\ell}$ and $\kappa_{q}$ separately
from $O(eeee)$ and $O(uudd)$ as follows:\begin{eqnarray}
\kappa_{\ell} & \leq & 8.6\frac{f}{{\rm TeV}},\label{eq:m_Todd_L}\\
\kappa_{q} & \leq & 37.1\frac{f}{{\rm TeV}}.\label{eq:m_Todd_q}\end{eqnarray}
However, $\kappa_{q}$ and $\kappa_{\ell}$ are correlated by the
$O(eedd)$ which leads to\begin{equation}
\frac{\kappa_{\ell}^{2}\kappa_{q}^{2}}{\kappa_{\ell}^{2}-\kappa_{q}^{2}}{\rm ln}(\frac{\kappa_{\ell}}{\kappa_{q}})\,\leq\frac{128\pi^{3}f^{2}}{(26.4\,{\rm TeV})^{2}}\,.\label{eq:O_eedd}\end{equation}
Fig.\ \ref{fig:allowed-k} shows the correlation of Eq.~(\ref{eq:O_eedd})
for various values of $f$. The region below each curve is the allowed
parameter space of $\kappa_{\ell}$ and $\kappa_{q}$ for the corresponding
$f$. The constraint is tight for small $f$: when $f=500\,{\rm GeV}$,
large $\kappa_{q}$ prefers smaller $\kappa_{\ell}$ and vice versa,
for example, $\kappa_{q}>4$ requires $\kappa_{\ell}<1$. This constraint
becomes quite loose when $f$ becomes large.

\section{Phenomenology of the $W_{H}$ pair production at the LHC\label{sec:Collider-lhc}}

The production rate of $W_{H}^{+}W_{H}^{-}$ pair at the LHC is sizable,
but the detection for its signatures at the hadron collider was expected
to be challenging~\cite{Hubisz:2004ft,Belyaev:2006jh}. However,
in this work we will demonstrate that the LHC not only has a great
potential to discover the collider signature of the $W_{H}^{+}W_{H}^{-}$
pair production, but also has the capability to explore enormous parameter
space of $f$ and $\kappa$. Below we present a detailed study of
the LHC phenomenology. 

At the LHC, we demand the two $W_{H}$ bosons both decay leptonically
in order to avoid the huge QCD backgrounds. We further require the
two charged leptons in the final state having different lepton flavors.
Hence, the collider signature of the signal events is $e^{+}\mu^{-}\met$
(or $e^{-}\mu^{+}\met$), where the missing energy ($\met$) is originated
from two $A_{H}$'s and two neutrinos. For simplicity, we will present
the study of $e^{+}\mu^{-}\met$ signature throughout this paper,
but it is very straightforward to include the contribution of $e^{-}\mu^{+}\met$
mode as those two decay modes are identical\ %
\footnote{The mass difference between $e$ and $\mu$ can be safely ignored
in our study since we are dealing with new particles whose masses
are at the order of TeV.%
}. 

\begin{figure}
\includegraphics[scale=0.5]{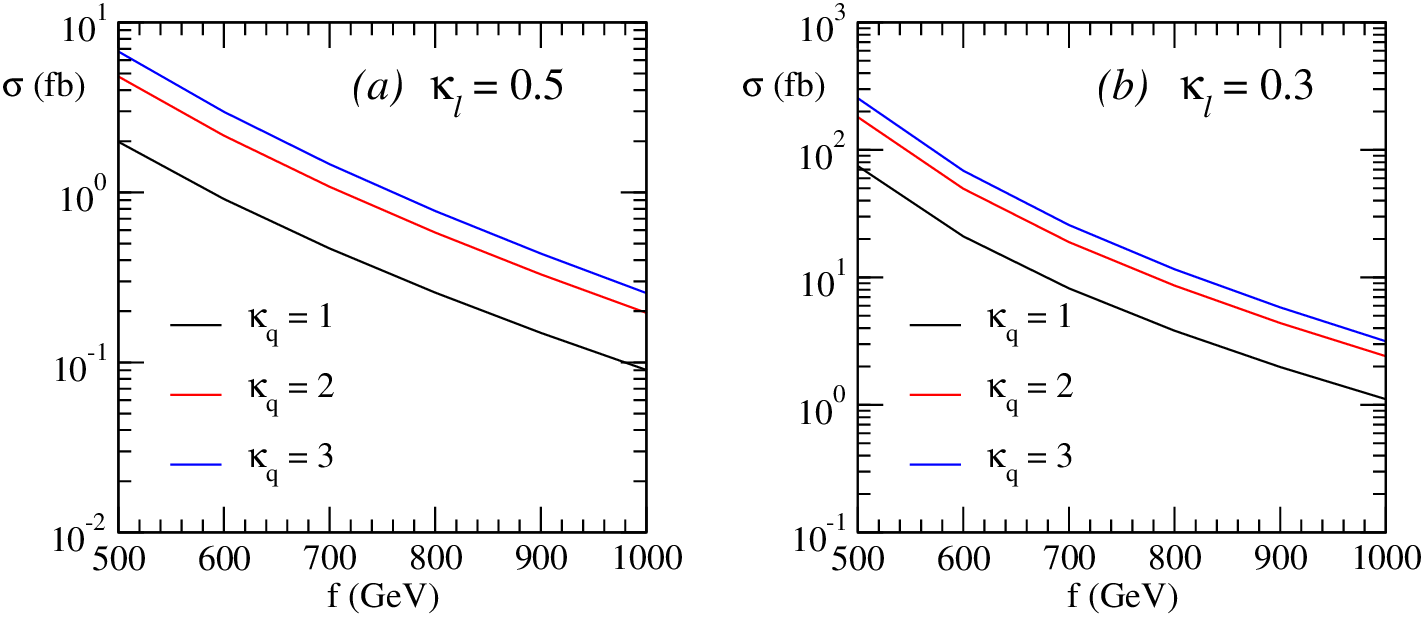}

\caption{The total cross section of $pp\to W_{H}^{+}W_{H}^{-}\to e^{+}\mu^{-}\met$
at the LHC.~\label{fig:rate-LHC}}
\end{figure}

When $W_{H}$ is the second lightest T-odd particle, i.e. $\kappa_{q}$
and $\kappa_{\ell}$ are both larger than 0.462, the signal events
only come from the following process \begin{equation}
pp\to W_{H}^{+}W_{H}^{-}\to A_{H}W^{+}(\to e^{+}\nu_{e})A_{H}W^{-}(\to\mu^{-}\bar{\nu}_{\mu}).\label{eq:signal}\end{equation}
However, when the T-odd leptons are lighter than $W_{H}$, i.e. $\kappa_{\ell}<0.462$,
the signal will mainly come from the process\begin{equation}
pp\to W_{H}^{+}W_{H}^{-}\to\ell_{1-}\ell_{2}\ell_{3-}\ell_{4}\to e^{+}\mu^{-}\nu_{e}\bar{\nu}_{\mu}A_{H}A_{H},\label{eq:signal2}\end{equation}
where $\ell_{i}=e,\,\mu,\,\nu_{e}$ or $\nu_{\mu}$. The total cross
sections of these two processes are shown in Fig.~\ref{fig:rate-LHC}
where the left plot is for the process in Eq.\ (\ref{eq:signal})
with $\kappa_{\ell}=0.5$ while the right plot is for the process
in Eq.\ (\ref{eq:signal2}) with $\kappa_{\ell}=0.3$. If $W_{H}$
is the second lightest T-odd particle, the signal will only come from
Eq.~(\ref{eq:signal}) since the $W_{H}$ can only decay to $WA_{H}$;
otherwise, the process in Eq.~(\ref{eq:signal2}) dominates. The
total rate of the signal events depends on the masses of $\ell_{-}$,
$q_{-}$ and $W_{H}$, and as shown in Fig.~\ref{fig:rate-LHC},
the total cross section is sizable when $f$ is small and $\kappa_{q}$
is large. This is because that the mass of T-odd gauge boson is light
and the destructive effect from t-channel and s-channel interference
term is small. 

The main intrinsic backgrounds come from the $W^{+}W^{-}$ and the
$ZW^{+}W^{-}$ continuum productions with the subsequent decays $W^{+}\to\ell^{+}\nu_{\ell}$,
$W^{-}\to\ell^{-}\bar{\nu}_{\ell}$ and $Z\to\nu\nu$\ %
\footnote{Generally speaking, we also need to consider the background from Higgs
boson decay into a $W$ boson pair, which is $gg\to H\to W^{+}W^{-}$
. The total rate depends on the mass of Higgs boson. For instance,
the total cross section is $\sim95\,{\rm fb}$ when the Higgs boson
is $120\,{\rm GeV}$, and $\sim230\,{\rm fb}$ when Higgs boson is
$170\,{\rm GeV}$. However, it can be completely suppressed by imposing
the kinematics cuts discussed later.%
}. There also exist other reducible backgrounds from the top quark
pair production and the $Wt$ associated production which can be highly
suppressed by vetoing the additional $b$-jet from the top quark decay
with large transverse momentum or in the central rapidity region.
The vetoing efficiency is so large, about $99.9\%$ for the $t\bar{t}$
background and $99.6\%$ for the $Wt$ background, that we only need
to consider the intrinsic backgrounds in this study. The total cross
section of the $W^{+}W^{-}$ pair production background is about $0.865\,{\rm pb}$
while the other intrinsic background from $W^{+}W^{-}Z$ is negligible
($\sim0.08\,{\rm fb}$). These cross sections already include the
decay branching ratios of $W\to\ell\nu$ and $Z\to\nu\nu$. Below,
we just consider the $W^{+}W^{-}$ pair production as the background
at the LHC.

\begin{figure}
\includegraphics[clip,scale=0.55]{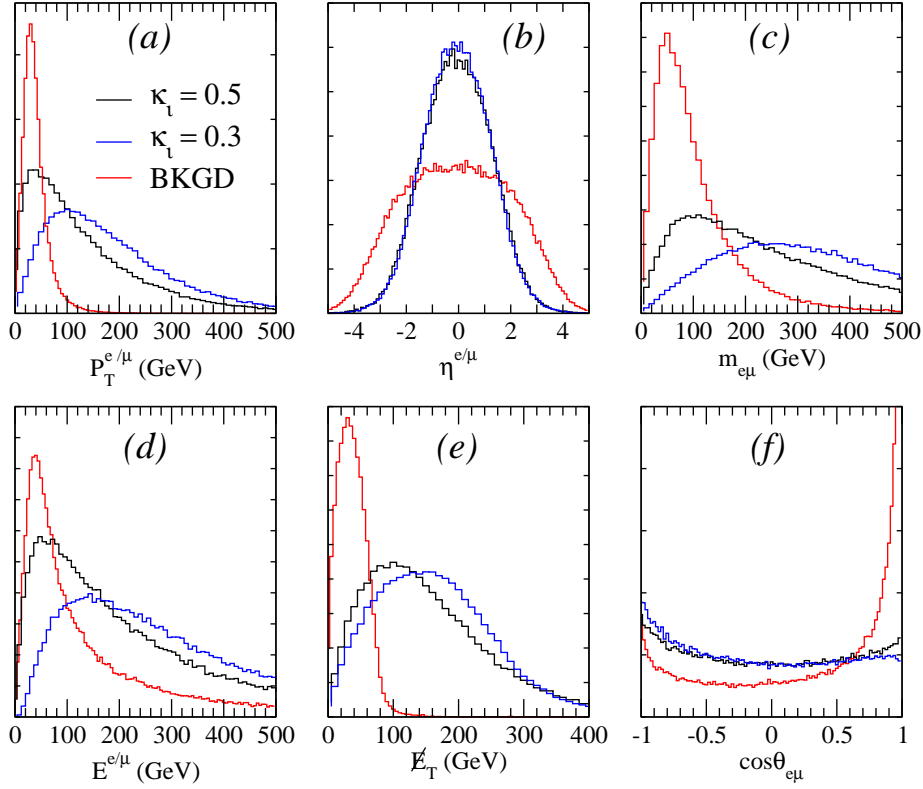}

\caption{Transverse momentum of $e^{+}/\mu^{-}$ ($p_{T}^{e/\mu}$), rapidity
of $e^{+}/\mu^{-}$ ($\eta^{e/\mu}$), invariant mass of $e^{+}$
and $\mu^{-}$ ($m_{e\mu}$), energy of $e^{+}/\mu^{-}$ ($E^{e/\mu})$,
missing transverse momentum ($\met$), cosine of the opening angle
between $e^{+}$ and $\mu^{-}$($\cos\theta_{e\mu}$) distributions
for $\kappa_{q}=1$ and $f=700\,{\rm GeV}$. All curves are normalized
by their total cross sections.~\label{fig:no-cut}}
\end{figure}

Kinematics of the signal events is distinctively different from that
of background events. As to be shown later, these differences can
be used to significantly suppress the background and enhance the ratio
of signal to background ($S/B$). For illustration, we show normalized
distributions of various kinematics observables of the signal and
background events in Fig.~\ref{fig:no-cut}: transverse momentum
$(p_{T}^{e/\mu})$, rapidity $(\eta^{e/\mu})$, energy $(E^{e/\mu})$
of charged leptons, invariant mass of two charged leptons ($m_{e\mu}$),
missing transverse momentum ($\met$) and cosine of the opening angle
between two charged leptons ($\cos\theta_{e\mu}$). The curves labelled
by $\kappa_{\ell}=0.5$ and $\kappa_{\ell}=0.3$ correspond to the
signals described in Eq.\ (\ref{eq:signal}) and Eq.\ (\ref{eq:signal2}),
respectively. A few interesting points are summarized below: 

\begin{itemize}
\item Compared to the background, the typical feature of the signal events
is that the final state particles are more energetic, cf. Fig.\ \ref{fig:no-cut}(a),
(c), (d), (e). 
\item As the decay products of heavy $W_{H}$ bosons, the two charged leptons
mainly appear in the central region, cf. Fig.\ \ref{fig:no-cut}(b),
because $W_{H}$ is hardly boosted.
\item We also note that, unlike the background, two charged leptons of the
signal do not exhibit strong correlations, see the nearly flat behavior
in the $\cos\theta_{e\mu}$ distribution. It can be understand as
follows. Since $m_{W_{H}}$ is much larger than $m_{W}$ and $m_{A_{H}}$,
$W$ and $A_{H}$ will be predominately in the longitudinal polarization
state, i.e. behaving as scalars. Thus, the spin correlation between
$e^{+}$ and $\mu^{-}$ is lost, which results in a flat distribution.
On the contrary, the two charged leptons in the SM background are
highly correlated. 
\item The signal distributions change a lot when varying the value of $\kappa_{\ell}$.
In particular, for a Small $\kappa_{\ell}$, i.e. $\kappa_{\ell}=0.3$,
the peak positions of the $p_{T}^{e/\mu}$, $m_{e\mu}$, $E_{e}^{e/\mu}$
and $\met$ distributions are shifted to the large value region when
compared to those of large $\kappa_{\ell}$, i.e. $\kappa_{\ell}=0.5$.
This is due to the fact that for a small $\kappa_{\ell}$, the charged
leptons ($e^{+}$ and $\mu^{-}$) or the neutrinos ($\nu_{e}$ and
$\bar{\nu}_{\mu}$) are directly generated from the $W_{H}$ boson
decay, e.g. $W_{H}^{+}\to e^{+}\nu_{e_{-}}$ or $W_{H}^{+}\to\nu_{e}e_{-}^{+}$,
and therefore are more energetic.
\end{itemize}
In order to mimic the detector, we require $p_{T}^{e/\mu}$ and $\eta^{e/\mu}$
to satisfy the following basic cuts:\begin{eqnarray}
 &  & p_{T}^{e}>20.0\,{\rm GeV}\,,\, p_{T}^{\mu}>20.0\,{\rm GeV}\,,\nonumber \\
 &  & \,|\eta^{e}|<2.0\,\,\,\,\,\,\,\,\,\,\,\,\,,\,|\eta^{\mu}|<2.0\,.\label{eq:basic-cut}\end{eqnarray}
Furthermore, taking advantage of the differences between the kinematics
of the signal and background events, we impose the following \emph{optimal
cuts} to extract the signal out of the SM background,\begin{equation}
\met>175\,{\rm GeV}\,,\,\cos\theta_{e\mu}<0.6\,.\label{eq:cut}\end{equation}

\begin{figure}
\includegraphics[clip,scale=0.6]{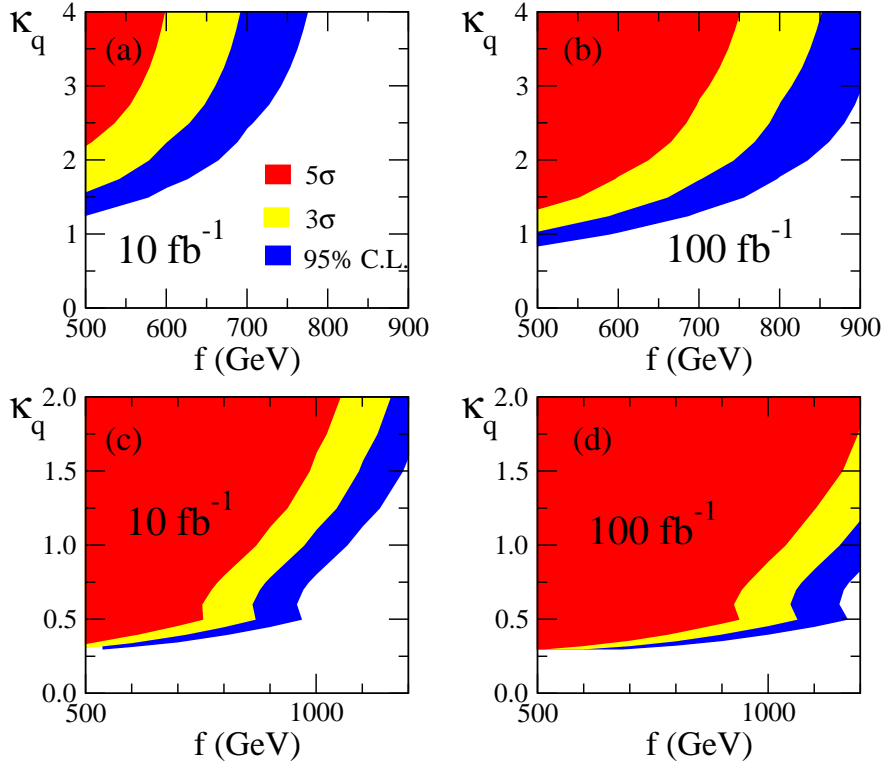}

\caption{Statistical significance contour of signature of $pp\to W_{H}^{+}W_{H}^{-}\to\ell^{+}\ell^{\prime-}\nu_{\ell}\bar{\nu}_{\ell^{\prime}}A_{H}A_{H}$
in the plane of $\kappa_{q}$ and $f$ at the LHC. The upper two plots
are for $\kappa_{\ell}=0.5$ while the lower two are for $\kappa_{\ell}=0.3$.~\label{fig:contour} }
\end{figure}

After imposing the optimal cuts, the main background from the $W^{+}W^{-}$
pair production can be suppressed by more than $99\%$ and gives rise
to 18 background events for $\mathcal{L}=10\,{\rm fb}^{-1}$ while
192 events for $\mathcal{L}=100\,{\rm fb}^{-1}$, where $\mathcal{L}$
denotes the integrated luminosity. These background rates include
both $e^{+}\mu^{-}$ and $e^{-}\mu^{+}$ modes. In Fig.~\ref{fig:contour}
we present the $5\sigma$, $3\sigma$ statistical significance and
$95\%$ confidence level (C.L.) for $\kappa_{\ell}=0.5$ (top raw)
and $\kappa_{\ell}=0.3$ (bottom raw). For $\kappa_{\ell}=0.5$, the
$W_{H}$ boson is the second lightest T-odd particle and the signal
events come from Eq.~(\ref{eq:signal}) only. When $f$ is $500\,{\rm GeV}$,
the signal can reach more than $3\sigma$ statistical significance
for $\kappa_{q}\gtrsim1.5$ with $\mathcal{L}=10\,{\rm fb}^{-1}$
and $\kappa_{q}\gtrsim1$ with $\mathcal{L}=100\,{\rm fb}^{-1}$,
respectively. Furthermore, the $f$ can be probed up to about $770\,{\rm GeV}$
with $\mathcal{L}=10\,{\rm fb}^{-1}$ and $950\,{\rm GeV}$ with $\mathcal{L}=100\,{\rm fb}^{-1}$,
respectively, at the $95\%$ C.L.. On the other hand, for $\kappa_{\ell}=0.3$,
the T-odd leptons are lighter than $W_{H}$ and the signal events
predominantly come from Eq.~(\ref{eq:signal2}) due to the large
decay branching ratios. In this case, one can probe more parameter
space of the LHT, cf. Fig.\ \ref{fig:contour}(c)\ and\ (d). For
example, assuming $\kappa_{q}=1$, one can probe $f$ up to $900\,{\rm GeV}$
with $\mathcal{L}=10\,{\rm fb}^{-1}$ and $1050\,{\rm GeV}$ with
$\mathcal{L}=100\,{\rm fb}^{-1}$, respectively, at the $5\sigma$
level.

\begin{figure}
\includegraphics[clip,scale=0.65]{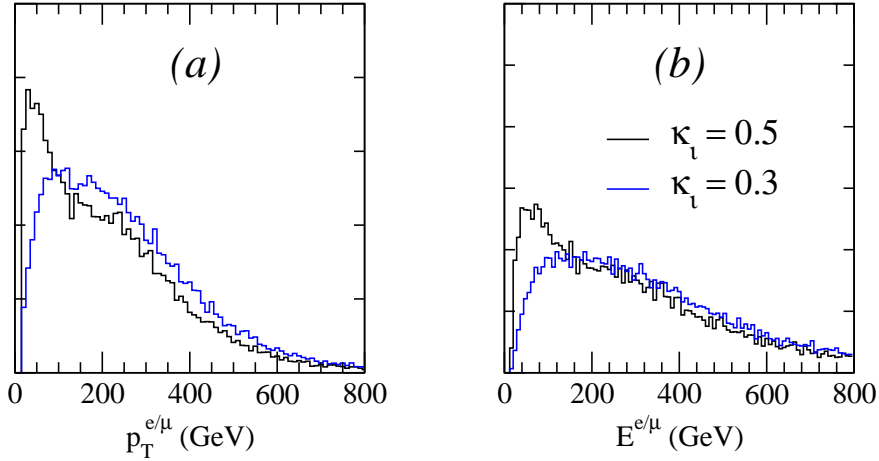}

\caption{Normalized distributions of $p_{T}^{e/\mu}$ and $E^{e/\mu}$ for
$f=700\,{\rm GeV}$ and $\kappa_{q}=1$ for $pp\to W_{H}^{+}W_{H}^{-}\to e^{+}\mu^{-}\nu_{e}\bar{\nu}_{\mu}A_{H}A_{H}$
process after imposing the kinematics cuts given in Eq.~(\ref{eq:cut})
at the LHC.~\label{fig:kin_cut}}
\end{figure}

As shown above, it is very promising to use the $e\mu+\met$ signature
to detect the $W_{H}W_{H}$ pair production at the LHC. But such a
signature can originate from two processes, either Eq.\ (\ref{eq:signal})
or Eq.\ (\ref{eq:signal2}), depending on the value of $\kappa_{\ell}$.
Therefore, one immediate task after observing such a signature is
to determine from which process it comes. It turns out that this question
can be easily answered by the $p_{T}^{e/\mu}$ and $E^{e/\mu}$ distributions,
cf. Fig.~\ref{fig:kin_cut} where we have imposed the optimal cuts.
In case of $\kappa_{\ell}=0.3$, the charged lepton is directly emitted
from the T-odd gauge boson decay, therefore its transverse momentum
is typically larger than the one of the charged lepton emitted form
the $W$-boson decay, i.e. $\kappa_{\ell}=0.5$. Same argument also
works for the energy distributions. Hence, one can fit the observed
$p_{T}^{e/\mu}$ and $E^{e/\mu}$ distributions to the LHT predictions
to measure $\kappa_{\ell}$, though $\kappa_{q}$, which merely change
the normalization of both distributions, remains unknown. 

\section{Phenomenology of the $W_{H}$ pair production at the LC\ \label{sec:Collider-ilc}}

Compared to the LHC, the LC does not have sufficient energy to produce
very heavy $W_{H}$ bosons. For example, the LC can only probe the
$W_{H}$ boson mass up to $500\,{\rm GeV}$, which corresponds to
$f\simeq750\,{\rm GeV}$. However, the LC provides a much cleaner
experimental environment (no QCD backgrounds) which is perfect for
precision measurements. As mentioned before, because of suffering
from the extremely huge QCD backgrounds, one has to use the leptonic
decay mode for the $W_{H}$ boson search at the LHC. One can observe
a deviation from the SM prediction, but one cannot determine the mass
or spin of the $W_{H}$ boson due to the four missing particles (two
$A_{H}$'s and two neutrinos) in the final state. In this section
we preform a comprehensive study of the $W_{H}$ pair production at
the LC and address on the following questions: 

\begin{itemize}
\item Can one determine the masses of $W_{H}$ and $A_{H}$?
\item Can we reconstruct the kinematics of the missing particle $A_{H}$?
\item Can we measure the spin of $W_{H}$?
\end{itemize}
As to be shown later, all these questions can be easily answered at
the LC with the help of the known center-of-mass (c.m.) energy. 

\begin{figure}
\includegraphics[clip,scale=0.5]{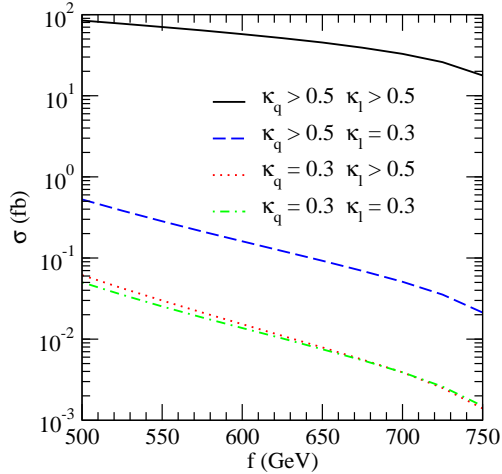}

\caption{Total cross section for $e^{+}e^{-}\to W_{H}^{+}W_{H}^{-}\to A_{H}A_{H}jjjj$
at the LC.~\label{fig:4jets}}
\end{figure}

At the LC, we are able to search the $W_{H}$ boson using its hadronic
decay mode $W_{H}\to A_{H}W\to A_{H}jj$. Below, we consider the following
signal process\begin{equation}
e^{+}e^{-}\to W_{H}^{+}W_{H}^{-}\to W^{+}(\to jj)W^{-}(\to jj)A_{H}A_{H},\label{eq:signal-process-ilc}\end{equation}
which gives rise to a collider signature of four isolated jets associated
with large missing energy originated from the two undetectable $A_{H}$
bosons in the final state. The main intrinsic background is from the
process $e^{+}e^{-}\to W^{+}W^{-}Z\to jjjj\nu\bar{\nu}$ whose cross
section is about $5.6\,{\rm fb}$. In Fig.~\ref{fig:4jets}, we show
the cross section of the signal process given in Eq.\ (\ref{eq:signal-process-ilc})
at the LC. The total cross section relies on how large the decay branching
ratio of the $W_{H}\to WA_{H}$ mode is: (1) when both $\kappa_{\ell}$
and $\kappa_{\ell}$ are large, $Br(W_{H}\to WA_{H})=1$ which leads
to a large cross section, see the black (solid) curve; (2) when either
$\kappa_{q}$ or $\kappa_{\ell}$ is small, $Br(W_{H}\to WA_{H})$
is highly suppressed, so the total cross section becomes small, see
the blue (dashed), the red (dotted) and the green (dot-dashed) curves.
In this work we focus our attention on the first case, i.e. large
$\kappa_{q}$ and $\kappa_{\ell}$, in which $W_{H}$ is the second
lightest T-odd particle. Since the cross section of the signal process
is much higher than the $WWZ$ background, it is not difficult to
disentangle the signal from the background. Therefore, only the basic
kinematics cuts, but no further hard cuts, are applied to select the
event in the following study. For comparison, we also present the
background distributions. 

When either $\kappa_{q}$ or $\kappa_{\ell}$ is small, one has to
consider other decay modes to search the $W_{H}$ boson. For example,
when $\kappa_{q}=0.3$, the T-odd quark is lighter than the $W_{H}$
boson. One thus can use the following process\begin{equation}
e^{+}e^{-}\to W_{H}^{+}W_{H}^{-}\to qq_{-}^{\prime}qq_{-}^{\prime}\to qqqqA_{H}A_{H}\label{eq:signal-lc-2}\end{equation}
 to search the $W_{H}$ boson. Searching the $W_{H}$ boson in this
channel is very interesting but certainly beyond the scope of this
work. Detailed study of this channel will be presented elsewhere.

\subsection{Mass measurement of $W_{H}$}

In order to simulate the detector acceptance, we require the transverse
momentum ($p_{T}^{j}$) and rapidity ($\eta^{j}$) of all the final
state jets to satisfy the following basic cuts\[
p_{T}^{j}>15\,{\rm GeV},\qquad\left|\eta^{j}\right|<3\,.\]
We also demand that the four jets are resolvable as separated objects,
i.e. requiring the separation in $\Delta R\equiv\sqrt{(\delta\eta)^{2}+(\delta\phi)^{2}}$
between any two jets to be larger than 0.4, where $\delta\eta$ and
$\delta\phi$ denotes the separation in the rapidity and azimuthal
angles, respectively. In order to reconstruct the two $W$ bosons,
one need to isolate the four jets coming from the $W$ boson decay.
Unfortunately, one cannot tell the jets apart experimentally because
the information of quark's charge and flavor is lost in the hadronization
of the light quarks. In order to measure $m_{W_{H}}$, one needs to
reconstruct the two $W$ bosons, i.e. finding out which two jets come
from which $W$ boson. In this study we use the $W$ boson mass as
a constraint to reconstruct two $W$ bosons:

\begin{itemize}
\item In order to identify the jets, we order the four jets by their transverse
momentum, \begin{equation}
p_{T}^{j_{1}}\ge p_{T}^{j_{2}}\ge p_{T}^{j_{3}}\ge p_{T}^{j_{4}}.\end{equation}

\item We loop over all combinations of the four jets, i.e. ($j_{1}j_{2}$,
$j_{3}j_{4}$), ($j_{1}j_{3}$, $j_{2}j_{4}$) and ($j_{1}j_{4}$,
$j_{2}j_{3}$), and calculate the invariant masses of the reconstructed
$W$ bosons. We then calculate the deviations from the true $W$ boson
mass ($m_{W}$) for each combination,\begin{equation}
\Delta=\sqrt{\left(m_{1}(jj)-m_{W}\right)^{2}+\left(m_{2}(jj)-m_{W}\right)^{2}},\label{eq:delta-wmass}\end{equation}
 and select the combination giving rise to the minimal deviations
to reconstruct the $W$ bosons. Although the efficiency of the $W$
boson reconstruction procedure is very high ($\sim99.1\%$), we cannot
distinguish the two reconstructed $W$ bosons because the charge information
is lost. But as to be shown below, we do not need the information
of the $W$ boson charge to determine the mass and spin of $W_{H}$.
Just for bookmark we denote the $W$ boson consisting the highest
$p_{T}$ jet as $W_{1}$ while the other $W$ boson as $W_{2}$.
\end{itemize}
\begin{figure}
\includegraphics[clip,scale=0.5]{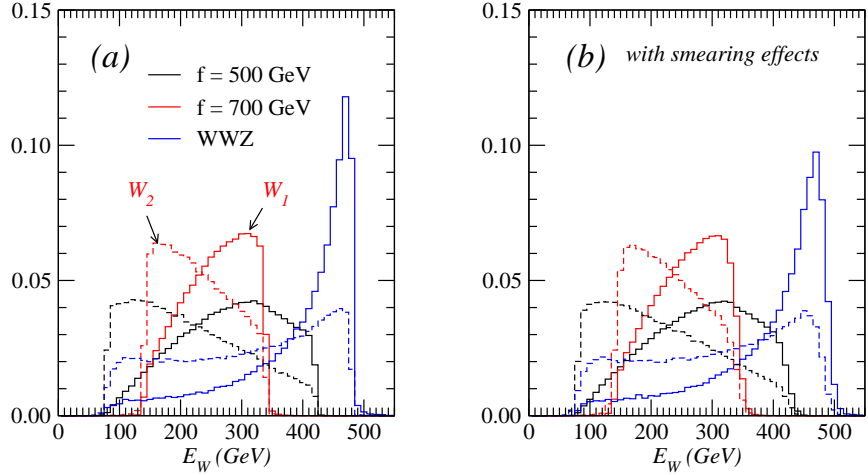}

\caption{Normalized energy distributions of the reconstructed $W$ bosons
for $\kappa_{q}=1$ at the LC.~\label{fig:EW-ilc}}
\end{figure}

In Fig.\ \ref{fig:EW-ilc}, we present the energy distributions of
the reconstructed $W$ bosons ($E_{W}$) where the energy of $W_{1}$
($E_{W_{1}}$) peaks in the large energy region while the energy of
$W_{2}$ ($E_{W_{2}}$) in the small energy region. The asymmetry
between $W_{1}$ and $W_{2}$ is due to our requirement that the $W_{1}$
boson includes the leading-$p_{T}$ jet. Since the $A_{H}$ bosons
are massive, the $E_{W}$ distributions exhibit sharp drops in both
small and large energy regions, which can be used to measure the masses
of $W_{H}$ and $A_{H}$~\cite{Kong:2007uu}. The ending points of
the energy distribution of the $W$ boson are given by\begin{equation}
E_{\pm}=\gamma\left(E_{W}^{\star}\pm\beta\, p_{W}^{\star}\right),\label{eq:eplusmin}\end{equation}
where $\beta=\sqrt{1-4m_{W_{H}}^{2}/s}$, $\gamma=1/\sqrt{1-\beta^{2}}$
and $E_{W}^{\star}$($p_{W}^{\star}$) is the energy (momentum) magnitude
of the $W$ boson in the rest frame of $W_{H}$,\begin{eqnarray}
E_{W}^{\star} & = & \frac{m_{W_{H}}^{2}-m_{A_{H}}^{2}+m_{W}^{2}}{2m_{W_{H}}},\label{eq:ew}\\
p_{W}^{\star} & = & \frac{\sqrt{\left[m_{W_{H}}^{2}-\left(m_{A_{H}}+m_{W}\right)^{2}\right]\left[m_{W_{H}}^{2}-\left(m_{A_{H}}-m_{W}\right)^{2}\right]}}{2m_{W_{H}}}.\label{eq:pw}\end{eqnarray}
From $E_{\pm}$ we can derive $m_{W_{H}}$ and $m_{A_{H}}$ as follows:\begin{eqnarray}
m_{W_{H}} & = & \sqrt{\frac{s}{2}}\frac{\sqrt{E_{+}E_{-}}}{E_{+}+E_{-}}\sqrt{1+\frac{m_{W}^{2}}{E_{+}E_{-}}+\sqrt{\left(1-\frac{m_{W}^{2}}{E_{+}^{2}}\right)\left(1-\frac{m_{W}^{2}}{E_{-}^{2}}\right)}},\label{eq:xmwh}\\
m_{A_{H}} & = & m_{W_{H}}\sqrt{1-\frac{2\left(E_{+}+E_{-}\right)}{\sqrt{s}}+\frac{m_{W}^{2}}{m_{W_{H}}^{2}}}.\label{eq:xmah}\end{eqnarray}
In this study, we choose two sample points: (1) $m_{W_{H}}=320\,{\rm GeV}$
and $m_{A_{H}}=66\,{\rm GeV}$ for $f=500\,{\rm GeV}$; (2) $m_{W_{H}}=450\,{\rm GeV}$
and $m_{A_{H}}=101\,{\rm GeV}$ for $f=700\,{\rm GeV}$. Hence, for
the former sample point, $E_{+}=426\,{\rm GeV}$ and $E_{-}=85\,{\rm GeV}$,
while for the latter sample point, $E_{+}=345\,{\rm GeV}$ and $E_{-}=146\,{\rm GeV}$.
The small tails of the lower and higher ending points are due to the
width effects of $W_{H}$ and $W$. After reading out the ending points
from the $E_{W}$ distribution, one can determine $m_{W_{H}}$ and
$m_{A_{H}}$ from Eqs.\ (\ref{eq:xmwh})\ and\ (\ref{eq:xmah}).
The accuracy of this method highly depends on how well one can reconstruct
the $W$ boson momentum and how well one can determine the ending
points. Furthermore, the collider detection is not perfect. In order
to mimic the finite detection efficiency of the detector, we smear
the momenta of all the final state jets by a Gaussian distribution
with \begin{equation}
\frac{\Delta E}{E}=\frac{50\%}{\sqrt{E}},\label{eq:smearing}\end{equation}
where $E$ is the energy of the observed parton and the resolution
of the energy measurement is assumed to be $50\%\sqrt{E}$. The $E_{W}$
distributions after energy smearing are shown in Fig.\ \ref{fig:EW-ilc}(b).
We note that the shapes of the distributions of both signal and background
are changed slightly, but the positions of the ending points remain
almost the same,  which lead to $4\%$ and $8\%$ error in the mass
measurements of $W_{H}$ and $A_{H}$ for $f=700$ GeV, respectively.

\subsection{Spin correlations}

Although one can derive the $W_{H}$ mass by using $E_{+}$ and $E_{-}$
from the $E_{W}$ distributions, one still needs to verify that such
a signal indeed comes from the LHT and not from other new physics
models. For example, the minimal supersymmetric extension of the standard
model (MSSM) with R-parity can also have exactly the same collider
signature ($4j+\met$) from the process \[
e^{+}e^{-}\to\widetilde{W}^{+}\widetilde{W}^{-}\to\tilde{\gamma}\tilde{\gamma}W^{+}(\to jj)W^{-}(\to jj),\]
where the photino ($\tilde{\gamma}$) is assumed to be the lightest
SUSY particle which plays as the dark matter candidate. Obviously,
examining the kinematics distributions is not sufficient to discriminate
the LHT from the MSSM. Below we will show that the spin correlation
between the $W$ boson and its mother particle is a good tool to tell
these two models apart. Taking advantage of the known c.m. energy
of the LC, one can reconstruct the kinematics of the two missing $A_{H}$
bosons and in turn study the spin correlation effects for model discrimination.
Details of the event reconstruction are shown in the Appendix. Below,
we only present our results of the phenomenological study. 

\begin{table}

\caption{Efficiencies of the $A_{H}$ reconstruction after requiring $\mathbb{C}^{2}>0$.\label{tab:AH-reconstruction-efficicency}}

\begin{tabular}{cccccccccccc}
\hline 
$f$ (GeV)&
&
\multicolumn{2}{c}{input (GeV)}&
&
\multicolumn{3}{c}{no smearing}&
&
\multicolumn{3}{c}{with smearing}\tabularnewline
\hline 
&
\ \ &
$m_{W_{H}}$&
$m_{A_{H}}$&
$\qquad$&
signal&
$\quad$&
BKGD&
$\qquad$&
signal&
$\quad$&
BKGD\tabularnewline
\hline 
500&
&
317&
66&
&
$87\%$&
&
$0.5\%$&
&
$80\%$&
&
$1.4\%$\tabularnewline
\hline 
600&
&
384&
84&
&
$90\%$&
&
$0.3\%$&
&
$82\%$&
&
$0.7\%$\tabularnewline
\hline 
700&
&
450&
101&
&
$89\%$&
&
$0.1\%$&
&
$79\%$&
&
$0.3\%$\tabularnewline
\hline
\end{tabular}
\end{table}

After event reconstruction, we denote $A_{H1}$ as the reconstructed
$A_{H}$ boson associated with $W_{1}$ while $A_{H2}$ as the one
with $W_{2}$. The inequality $\mathbb{C}^{2}>0$ (cf. Eq.\ \ref{eq:appen-9}),
has to be satisfied in order to reconstruct the momentum of $A_{H}$'s.
Since $\mathbb{C}^{2}$ depends on $m_{W_{H}}$ and $m_{A_{H}}$,
inputting the correct masses of $W_{H}$ and $A_{H}$ will significantly
enhance the efficiency of the event reconstruction. Furthermore, it
is easy to show that the dependence of $\mathbb{C}^{2}$ upon $m_{W_{H}}$
is much stronger than the one upon $m_{A_{H}}$. Hence, if one inputs
the correct $m_{W_{H}}$, then one may reach the maximal reconstruction
efficiency. The reconstruction efficiencies are summarized in Table\ \ref{tab:AH-reconstruction-efficicency}
where we consider both cases of with and without detector smearing
effects. The detector effects reduce the efficiency of the signal
reconstruction about $10\%$ but increase the efficiency of the background
reconstruction by a factor $2\sim3$. 

\begin{figure}
\includegraphics[clip,scale=0.7]{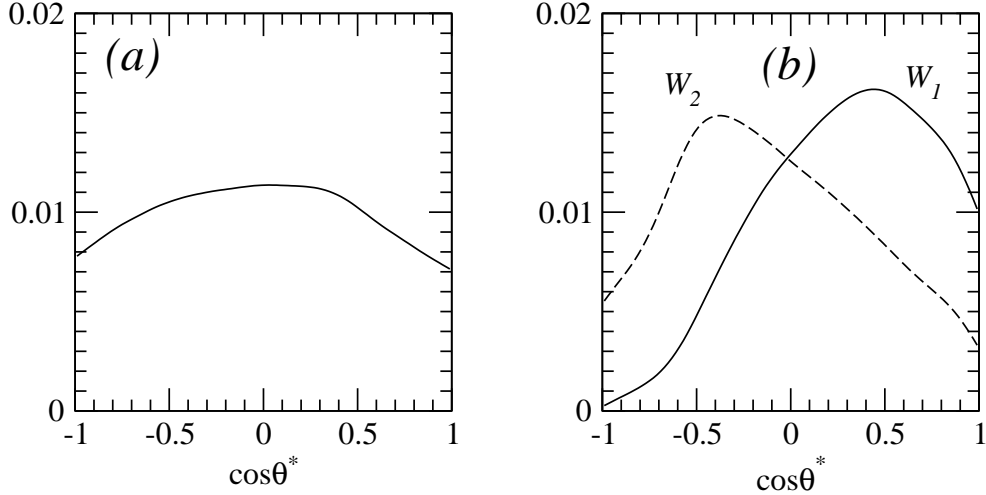}

\caption{Normalized distribution of $\cos\theta^{*}$, where $\theta^{*}$
is the angle between the $W$ boson and its mother particle $W_{H}$
in the rest frame of $W_{H}$ for $f=500$ GeV: (a) true distribution;
(b) after the $W$ boson reconstruction.~\label{fig:cthstar} }
\end{figure}

Using the known kinematics of the $A_{H}$ bosons, we can reconstruct
the momentum of the $W_{H}$ bosons. We then can plot the $\cos\theta^{*}$
distribution of the $W$ boson in Fig.\ \ref{fig:cthstar} where
$\theta^{\ast}$ is the angle between $W$ boson and $W_{H}$ boson
in the rest frame of $W_{H}$ boson. The left figure shows the true
$\cos\theta^{*}$ distribution where we assume all the particles in
the final state, including the $A_{H}$ bosons, are perfectly tagged.
The right figure shows the $\cos\theta^{*}$ distributions after the
$W$ boson reconstruction. The distributions can be understood as
follows. In the LHT, the decay products of the $W_{H}$ boson, $W$
and $A_{H}$, are highly boosted because $W_{H}$ is much heavier
than $A_{H}$ and $W$. Then the $A_{H}$ and $W$ bosons would be
predominately in the longitudinal polarization states. Therefore,
the decay of $W_{H}\to A_{H}W$ could be treated as a vector boson
decaying into two scalars. Due to the angular momentum conservation,
the spacial function of $A_{H}$ and $W_{H}$ would be dominated by
p-wave ($\sim\sin^{2}\theta^{\ast}$), as shown in Fig.~\ref{fig:cthstar}
(a). Duo to the $W$ boson reconstruction, cf. Fig\ \ref{fig:EW-ilc},
$W_{1}$, the $W$ boson containing the leading jet, moves parallel
with the $W_{H}$ and thus peaks in the forward direction while $W_{2}$
peaks in the backward direction. 

\begin{figure}
\includegraphics[clip,scale=0.7]{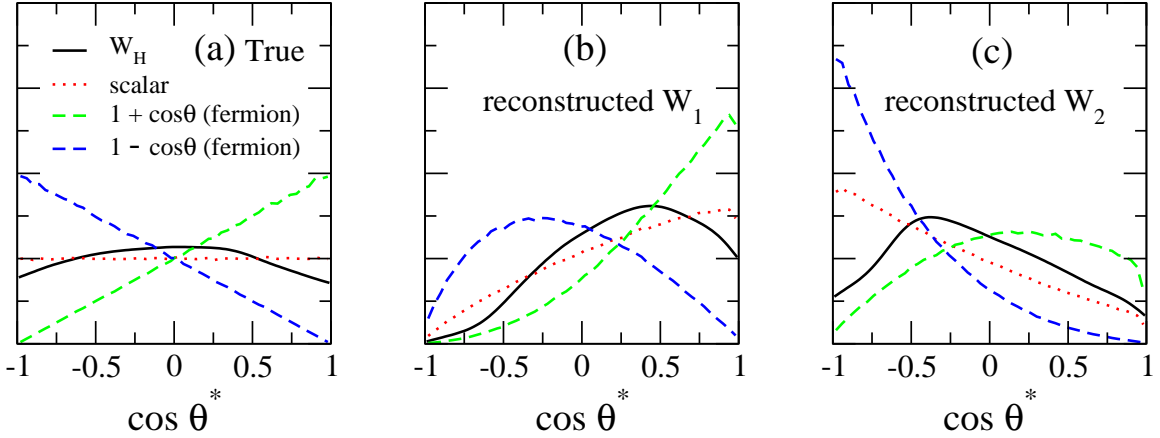}

\caption{Normalized $\cos\theta^{*}$ distributions for different spin particles:
(a) the true distribution while (b) and (c) are the distributions
after $W$ boson reconstruction.~\label{fig:cth_models}}
\end{figure}

How could we use this angular correlation to distinguish different
models? Let us consider the signature of $W^{+}W^{-}$+ $\met$ which
is generated by two heavy vector bosons in the LHT. That signature
could also be induced by many other new physics models: 

\begin{itemize}
\item It can come from the decays of a heavy scalar ($\Phi$) pair, e.g.
$e^{+}e^{-}\to\Phi\Phi\to W^{+}W^{-}+VV$, and the missing particle
($V$) must be a vector boson. Due to the scalar decay, the $\cos\theta^{*}$
distribution should be flat, cf. the red dotted curve in Fig.\ \ref{fig:cth_models}(a). 
\item It can also come from the decays of a heavy fermion ($\mathcal{F}$)
pair, e.g. $e^{+}e^{-}\to\mathcal{F}\mathcal{F}\to W^{+}W^{-}+\chi\chi$,
and the missing particle ($\chi$) must also be a fermion. It is well
know that the $\cos\theta^{*}$ distribution should be in the form
of $1-\cos\theta^{*}$, $1+\cos\theta^{*}$, or the combination of
them. Here we plot the first two distributions in Fig.\ \ref{fig:cth_models}(a),
cf. the blue dashed and green dashed curves\ %
\footnote{We note that the $\cos\theta^{*}$ distribution is flat if the heavy
fermion is unpolarized. It then is impossible to tell $\Phi$ and
$\mathcal{F}$ apart from the $\cos\theta^{*}$ distribution. However,
the distribution of the $W_{H}$ pair production in the LHT is still
distinguishable from those of $\Phi$ and $\mathcal{F}$.%
}.
\end{itemize}
The distinctive difference in the true $\cos\theta^{*}$ distributions
will be affected by the $W$ boson reconstruction, but the predictions
from different models are still distinguishable, cf. Fig.~\ref{fig:cth_models}(b)\ and\ (c).

\section{Conclusion~\label{sec:Conclusion}}

In this paper we study the collider phenomenologies of the $W_{H}$
pair production in the LHT at the LHC and the LC. The $W_{H}$ pair
production is of particular importance in the LHT because the mass
of $W_{H}$ is proportional to the symmetry breaking scale $f$. One
thus can unambiguously determine $f$ by measuring $m_{W_{H}}$. 

At the tree level, the $W_{H}$ boson pair can be produced either
via the $s$-channel process with the photon and $Z$ boson exchanged
or via the $t$-channel process with a T-odd fermion exchanged. The
total cross section highly relies on the mass of the T-odd fermion.
Although the $s$-channel and $t$-channel contributions are both
constructive, their interference effects are destructive. The total
cross section reaches the minimum when the $s$-channel and $t$-channel
contributions are comparable. Once being produced, the $W_{H}$ boson
will decay into a T-odd particle and a T-even SM particle. The decay
pattern of the $W_{H}$ boson is determined by the masses of other
new physics particles such as $A_{H}$, $\ell_{-}$ and $q_{-}$ (we
assume $A_{H}$ being the lightest T-odd particle):

\begin{enumerate}
\item If $W_{H}$ is the second lightest T-odd particle, it can only decay
into $A_{H}W$. 
\item If $W_{H}$ is heavier than $\ell_{-}$ and/or $q_{-}$, it will decay
into $\ell\ell_{-}$ and/or $qq_{-}$ as well as $A_{H}W$. 
\end{enumerate}
In this work we treat the $\kappa_{q}$ and $\kappa_{\ell}$ separately.
In order to avoid the FCNC problem, we further demand $\kappa_{q}$
and $\kappa_{\ell}$ being diagonal in the flavor space. 

To avoid the huge QCD background at the LHC, we require the $W_{H}$
boson decay leptonically. Hence, the signal events can come either
from the process in Eq.\ (\ref{eq:signal}) for $\kappa_{\ell}=0.5$
or from the process in Eq.\ (\ref{eq:signal2}) for $\kappa_{\ell}=0.3$.
We perform a Monte Carlo analysis of the signal process along with
the SM backgrounds and find that the $W_{H}$ boson decaying leptonically,
leading to a $\ell^{+}\ell^{\prime-}\met$ signature, is very promising
at the LHC. We apply the kinematical cuts in Eqs.~(\ref{eq:basic-cut},\ \ref{eq:cut})
and show the resulting significance contour in the plane of $\kappa_{q}$
and $\kappa_{\ell}$ in Fig.\ \ref{fig:contour}. We find that $f$
can be probed up to $750\,{\rm GeV}$ for $\kappa_{\ell}=0.5$ or
$1\,{\rm TeV}$ for $\kappa_{\ell}=0.3$ at the $5\sigma$ level with
an integrated luminosity of $100\,{\rm fb}^{-1}$. It is worth mentioning
that $f$ can be probed up to the same limits at the $95\%$ C.L.
even at low luminosity ($\mathcal{L}=10\,{\rm fb}^{-1}$) LHC operation.
Although the two processes given in Eqs.\ (\ref{eq:signal},\ \ref{eq:signal2})
give rise to the exactly same collider signature, they can be further
discriminated in the distributions of the transverse momentum and
energy of the final state charged leptons, see Fig.\ \ref{fig:kin_cut}.
However, the $W_{H}$ boson mass still cannot be determined at the
LHC due to the four missing particles in the final state. 

In order to determine the mass and spin of the $W_{H}$ boson, we
perform a Monte Carlo study of the $W_{H}$ pair production at the
LC. Owing to the clean background at the LC, we are able to search
the $W_{H}$ boson using its hadronic decay mode which leads to a
$4j+\met$ signature generated from Eq.\ (\ref{eq:signal-process-ilc}).
Due to the known center-of-mass energy at the LC, the masses of $W_{H}$
and $A_{H}$ can be determined from the ending points of the energy
distributions of the two reconstructed $W$ bosons. For example, one
can measure the mass of $W_{H}$ ($A_{H}$) within an error of $4\%$
($8\%$) for $f=700\,{\rm GeV}$, respectively, even after including
the detector smearing effects. Following the study of  the $W^{+}W^{-}$
pair production at the LEP\ \cite{Hagiwara:1986vm}, we present an
algorithm of reconstructing the kinematics of two undetectable $A_{H}$
bosons. It enables us to study the spin correlation between the $W$
boson and its mother particle ($W_{H}$) which is a powerful tool
to distinguish other new physics models from the LHT, as shown in
Fig.\ \ref{fig:cth_models}.

Combining the studies of the $W_{H}^{+}W_{H}^{-}$ pair production
at the LHC and LC, it is possible to determine or further constrain
the parameter $f$, $\kappa_{q}$ and $\kappa_{\ell}$. In order to
fix all the parameters of the LHT, direct search of other independent
channels, e.g. top quark partners (both T-odd and T-even) pair production
and T-odd fermions ( both leptons and quarks) pair production, must
be included in a systematic way. One then can compare all these independent
channels to check the consistence of the LHT. 

\begin{acknowledgments}
We thank Alexander Belyaev, Kazuhiro Tobe and C.-P. Yuan for useful
discussions. Q.-H. Cao is supported in part by the U.~S.~Department
of Energy under Grant No.~DE-FG03-94ER40837. C.-R. Chen is supported
in part by the U.S. National Science Foundation under award PHY-0555545.
\end{acknowledgments}
\appendix

\section{$A_{H}$ reconstruction at the LC\label{sec:Appendix-A}}

In this section we present an algorithm of determining the kinematics
of $A_{H}$ at the LC. This algorithm has been proposed in the study
of the $W$ boson at the LEP through the process $e^{+}e^{-}\to W^{+}W^{-}\to\ell^{+}\nu_{\ell}\ell^{\prime-}\bar{\nu}_{\ell^{\prime}}$~\cite{Hagiwara:1986vm}.
The difficulty is attributed to the existence of two missing particles
in the final state. The following kinematics analysis, presented below,
shows that the two unobserved momenta of $A_{H}$ bosons can be determined
from the reconstructed $W$ bosons up to a twofold discrete ambiguity,
in the limit where the $W$- and $W_{H}$-width are neglected. 

Here we consider the process \begin{equation}
e^{+}e^{-}\to AA',\,\,\, A\to BC,\,\, A'\to B'C'\end{equation}
where $A(A')$ is the mother particle while $B(B')$ and $C(C')$
are the decay products of the mother particles. Here we require $B(B')$
is observable while $C(C')$ undetectable. Furthermore, we assume\begin{equation}
m_{A}=m_{A'},\quad m_{C}=m_{C'}.\end{equation}
One of the advantage of the LC is the known center-of-mass energy
of the system. For example, the momentum of the incoming particles
are\begin{eqnarray}
p_{e^{+}} & = & \left(\begin{array}{cccc}
E_{t}, & 0, & 0, & E_{t}\end{array}\right),\\
p_{e^{-}} & = & \left(\begin{array}{cccc}
E_{t}, & 0, & 0, & -E_{t}\end{array}\right),\end{eqnarray}
where $E_{t}=\sqrt{S}/2$, where $\sqrt{S}$ is the total energy of
the linear collider.

From the momentum conservation, we obtain\begin{eqnarray}
E_{A}=E_{B}+E_{C}, &  & E_{A'}=E_{B'}+E_{C'},\label{eq:appen-1}\\
\vec{p}_{A}=\vec{p}_{B}+\vec{p}_{C}, &  & \vec{p}_{A'}=\vec{p}_{B'}+\vec{p}_{C'},\label{eq:appen-2}\end{eqnarray}
where $E_{i}(\vec{p}_{i})$ denotes the energy (three momentum) of
the particle $i$, respectively. At the LC, \begin{equation}
E_{A}=E_{A^{\prime}}=E_{t},\quad E_{C}=E_{t}-E_{B},\quad E_{C^{\prime}}=E_{t}-E_{B^{\prime}}.\end{equation}
From Eq.\ (\ref{eq:appen-2}) and the on-shell conditions of the
final state particles we obtain\begin{eqnarray}
2\vec{p}_{B}\cdot\vec{p}_{C} & = & E_{A}^{2}-m_{A}^{2}-\left(E_{B}^{2}-m_{B}^{2}\right)-\left(E_{C}^{2}-m_{C}^{2}\right),\label{eq:appen-3}\\
2\vec{p}_{B^{\prime}}\cdot\vec{p}_{C^{\prime}} & = & E_{A^{\prime}}^{2}-m_{A^{\prime}}^{2}-\left(E_{B^{\prime}}^{2}-m_{B^{\prime}}^{2}\right)-\left(E_{C^{\prime}}^{2}-m_{C^{\prime}}^{2}\right).\label{eq:appen-4}\end{eqnarray}
Using the momentum conservation \begin{equation}
\vec{p}_{B}+\vec{p}_{B^{\prime}}+\vec{p}_{C}+\vec{p}_{C^{\prime}}=0,\end{equation}
one obtains\begin{equation}
2\vec{p}_{B^{\prime}}\cdot\vec{p}_{C}=\left(E_{C^{\prime}}^{2}-m_{C^{\prime}}^{2}\right)-\left(E_{A^{\prime}}^{2}-m_{A^{\prime}}^{2}\right)-\left(E_{B^{\prime}}^{2}-m_{B^{\prime}}^{2}\right)-2\vec{p}_{B}\cdot\vec{p}_{B}.\label{eq:appen-5}\end{equation}
At last, the on-shell condition of particle $C$ gives us \begin{equation}
\left|\vec{p}_{C}\right|^{2}=E_{C}^{2}-m_{C}^{2}.\label{eq:appen-6}\end{equation}
Hence, one can determine $\vec{p}_{C}$ from Eqs.\ (\ref{eq:appen-3},\ \ref{eq:appen-5},\ \ref{eq:appen-6}).
We expand $\vec{p}_{C}$ in term of $\vec{p}_{B}$ and $\vec{p}_{B^{\prime}}$
as following\begin{equation}
\vec{p}_{C}=\mathbb{A}\vec{p}_{B}+\mathbb{B}\vec{p}_{B^{\prime}}+\mathbb{C}\vec{p}_{B}\times\vec{p}_{B^{\prime}}.\label{eq:appen-7}\end{equation}
Then one can derive $a$ and $b$ from Eqs.\ (\ref{eq:appen-3},\ \ref{eq:appen-5})
\begin{equation}
\left(\begin{array}{c}
\mathbb{A}\\
\mathbb{B}\end{array}\right)=\frac{1}{\left|\vec{p}_{B}\right|^{2}\left|\vec{p}_{B^{\prime}}\right|^{2}-\left(\vec{p}_{B}\cdot\vec{p}_{B^{\prime}}\right)^{2}}\left(\begin{array}{cc}
\left|\vec{p}_{B^{\prime}}\right|^{2} & -\vec{p}_{B}\cdot\vec{p}_{B^{\prime}}\\
-\vec{p}_{B}\cdot\vec{p}_{B^{\prime}} & \left|\vec{p}_{B}\right|^{2}\end{array}\right)\left(\begin{array}{c}
M\\
N\end{array}\right),\label{eq:appen-8}\end{equation}
where \begin{eqnarray}
M & \equiv & \frac{1}{2}\left[E_{A}^{2}-m_{A}^{2}-\left(E_{B}^{2}-m_{B}^{2}\right)-\left(E_{C}^{2}-m_{C}^{2}\right)\right],\\
N & \equiv & \frac{1}{2}\left[\left(E_{C^{\prime}}^{2}-m_{C^{\prime}}^{2}\right)-\left(E_{A^{\prime}}^{2}-m_{A^{\prime}}^{2}\right)-\left(E_{B^{\prime}}^{2}-m_{B^{\prime}}^{2}\right)-2\vec{p}_{B}\cdot\vec{p}_{B}\right].\end{eqnarray}
The remaining variable $\mathbb{C}$ is determined using Eq.\ (\ref{eq:appen-6}):\begin{equation}
\mathbb{C}^{2}=\frac{1}{\left|\vec{p}_{B}\times\vec{p}_{B^{\prime}}\right|^{2}}\left[E_{C}^{2}-m_{C}^{2}-\mathbb{A}^{2}\left|\vec{p}_{B}\right|^{2}-\mathbb{B}^{2}\left|\vec{p}_{B^{\prime}}\right|^{2}-2\mathbb{A}\mathbb{B}\vec{p}_{B}\cdot\vec{p}_{B^{\prime}}\right].\label{eq:appen-9}\end{equation}
The sign of $\mathbb{C}$ cannot be determined. This explicitly exhibits
a twofold discrete ambiguity. The inequality $\mathbb{C}^{2}>0$ is
expected to be violated only by finite $W$- and $W_{H}$-width effects.
Needless to say, using wrong $m_{C}$ and $m_{A}$ will lead to a
negative $\mathbb{C}^{2}$ which can serve to measure $m_{A}$ and
$m_{C}$ as mentioned earlier. In the exceptional case where the momenta
of particle $B$ and $B^{\prime}$ are parallel, one obtains a one-parameter
family of solution for which the azimuthal angle of $\vec{p}_{C}$
with respect to $\vec{p}_{B}$ is left undetermined. \bibliographystyle{apsrev}
\bibliography{reference}

\end{document}